\documentclass[fleqn,usenatbib]{mnras}

\usepackage{newtxtext,newtxmath}

\usepackage[T1]{fontenc}
\usepackage{ae,aecompl}

\usepackage{color}
\usepackage{lscape}
\usepackage{pdflscape}

\usepackage{breqn}

\usepackage{graphicx}	
\usepackage{amsmath}	
\usepackage{amssymb}	

\title[Lithium-strong KM dwarfs in GALAH]{The GALAH Survey: Lithium-strong KM dwarfs}

\author[M. {\v Z}erjal et al.]{
M. {\v Z}erjal,$^{1}$\thanks{E-mail: marusa.zerjal@anu.edu.au}
M. J. Ireland,$^{1}$
T. Nordlander,$^{1,2}$
J. Lin,$^{1}$
L. Casagrande,$^{1}$
J. Horner,$^{3}$
\newauthor 
G. De Silva$^{4,5}$
S. Martell,$^{6}$
K. {\v C}otar,$^7$
G. Traven,$^8$
T. Zwitter,$^7$
\newauthor 
and the GALAH Collaboration
\\
$^{1}$Research School of Astronomy \& Astrophysics, Australian National University, Canberra, ACT 2611, Australia\\
$^{2}$ARC Centre of Excellence for All Sky Astrophysics in 3 Dimensions (ASTRO 3D)\\
$^{3}$Centre for Astrophysics, University of Southern Queensland, Toowoomba, Qld 4350, Australia \\
$^{4}$Sydney Institute for Astronomy, School of Physics, A28, The University of Sydney, Sydney, NSW 2006, Australia \\
$^{5}$Australian Astronomical Observatory, 105 Delhi Rd, North Ryde, NSW 2113, Australia \\
$^{6}$School of Physics, UNSW, Sydney, NSW 2052, Australia \\
$^7$Faculty of Mathematics and Physics, University of Ljubljana, Jadranska 19, 1000 Ljubljana, Slovenia\\
$^8$Lund Observatory, Department of Astronomy and Theoretical Physics, Box 43, SE-221 00 Lund, Sweden
}

\date{Accepted XXX. Received YYY; in original form ZZZ}

\pubyear{2018}

\begin{document}
\label{firstpage}
\pagerange{\pageref{firstpage}--\pageref{lastpage}}
\maketitle

\begin{abstract}
Identifying and characterizing young stars in the Solar neighbourhood is essential to find and describe planets in the early stages of their evolution. This work seeks to identify nearby young stars showing a Lithium 6707.78{$\,$\AA} absorption line in the GALAH survey. A robust, data-driven approach is used to search for corresponding templates in the pool of 434,215 measured dwarf spectra in the survey. It enables a model-free search for best-matching spectral templates for all stars, including M dwarfs with strong molecular absorption bands. 
3147 stars have been found to have measurable Lithium: 1408 G and 892 K0-K5 dwarfs (EW(Li)$>$0.1{$\,$\AA}), 335 K5-K9 ($>$0.07{$\,$\AA}) and 512 M0$\sim$M4 dwarfs ($>$0.05{$\,$\AA}).  
Stars with such Lithium features are used to investigate the possibility of searching for young stars above the main sequence based merely on their parallaxes and broad-band photometry.
Selection of young stars above the main sequence is highly effective for M dwarfs, moderately effective for K dwarfs and ineffective for G dwarfs. Using a combination of the Lithium information and the complete 6D kinematics from {\it Gaia} and GALAH, 305 new candidate moving group members have been found, 123 of which belong to the Scorpius-Centaurus association, 36 to the Pleiades and 25 to the Hyades clusters.
\end{abstract}




\section{Introduction}
Over the past decade, a variety of surveys have greatly enhanced our knowledge of planets orbiting other stars. The {\it Kepler} space telescope \citep{2010Sci...327..977B} carried out the first large census of the Exoplanet Era, and revealed that planets on short-period orbits are ubiquitous \citep[e.g.][]{2013ApJS..204...24B,2015ApJS..217...31M}, whilst radial velocity surveys have begun to discover the first true Jupiter-analogues around Sun-like stars \citep[e.g.][]{2002ApJ...581.1375M,2012A&A...545A..55B,2014ApJ...783..103W,2016ApJ...819...28W}. In the coming years, it is widely anticipated that the number of known planets will increase dramatically, with NASA's Transiting Exoplanet Survey Satellite \citep[TESS; launched in April 2018,][]{2015JATIS...1a4003R} expected to deliver thousands of new planets in the next two years. Complementing the results of TESS, the {\it Gaia} mission may discover as many as 20,000 additional exoplanets by astrometry \citep{2014ApJ...797...14P}. 
Indeed the first exoplanet to be both directly imaged (including a spectrum and planet rotational velocity) and to have a measured mass recently had its mass measured using a combination of an early {\it Gaia} data release and HIPPARCOS \citep{2018arXiv180806257S}. 

\vspace{1em}

In take full advantage of these recent discoveries, and the plethora of new finds expected as a result of {\it TESS} and {\it Gaia}, it is imperative that we understand the basic characteristics of those stars which host them. Such knowledge is critical to our understanding of the formation and evolution of those planets. For example, there appears to be a lack of hot Jupiters around $\sim <$20$\,$Myr old stars \citep{2017AJ....154..224R}, hinting that hot Jupiters may not form by disk-driven migration. This tentative result requires age knowledge of a much larger sample of stars, both for detections and null results with TESS. Interpreting the luminosity of giant planets detected astrometrically requires robust age indicators, especially for young stars where planetary system evolution is most rapid and there is a chance of directly detecting thermal emission from giant planets.

\begin{figure*}
\includegraphics[width=\linewidth]{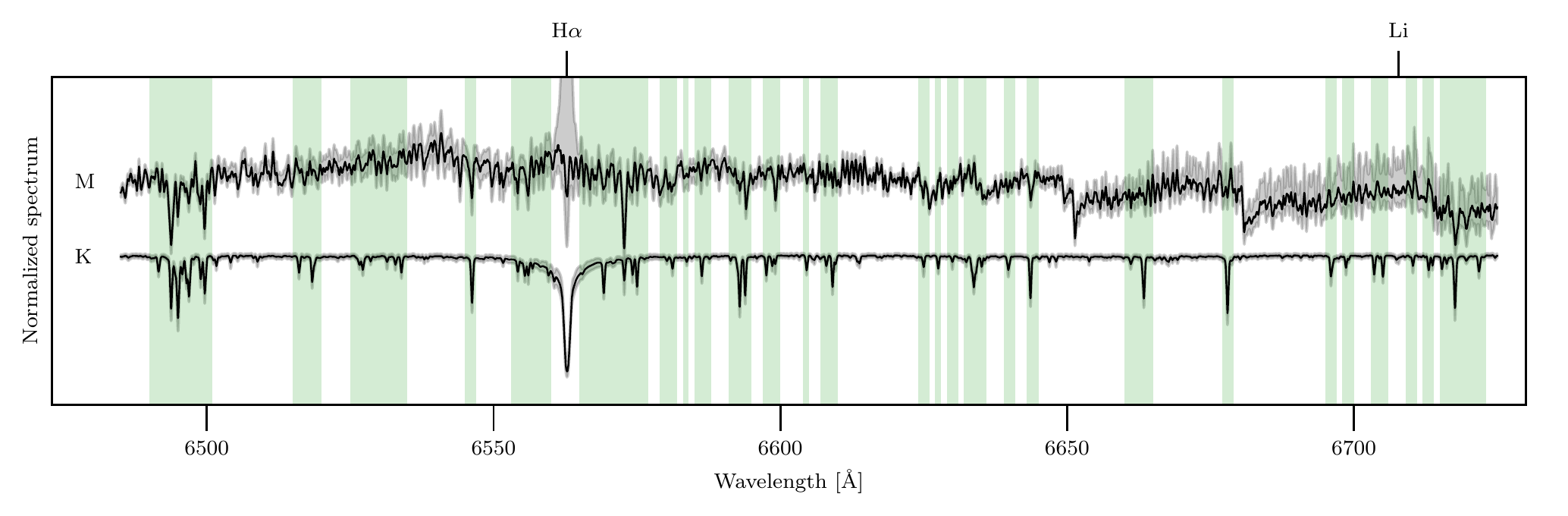}
\caption{The averaged early K and early M spectra in GALAH with deviations (grey areas) due to parameter variations. M dwarfs often show strong H$\alpha$ emission and higher variation of pseudo-continuum above 6650{$\,$\AA} among the sample. Wavelength regions used in the nearest neighbour search are shown by the green bands.}
\label{fig.bands}
\end{figure*}


The importance of such accurate characterisation can be seen from studies of the directly imaged objects orbiting the star HR 8799 \citep{2008Sci...322.1348M,2010Natur.468.1080M,2016AJ....152...28K,2017A&A...598A..83W}. The precise mass and nature of those planets is highly dependent on the accurate parameterisation of the star, the results of which could have a significant impact on the stability of the planets, and the mass of the associated debris disks in that system \citep[e.g.][]{2010IJAsB...9..259M,2012ApJ...761...57B,2014MNRAS.440.3140G,2016A&A...592A.147G,2016MNRAS.463..191C}. Over the decade since the HR8799 system was discovered, a better understanding of the nature and age of the host star has resulted in greater surety that the detected objects are truly planetary in nature.
The search for young stars near the Sun, especially those younger than 100$\,$Myr, is thus essential to understand planet formation conditions and their early evolution.


During the early stages in their lives, stars tend to share a common motion with their siblings, and to lie in the same region of space.
Recently, \citet[][hereafter G18]{2018ApJ...856...23G} identified 1406 members of 27 young stellar associations and moving groups within 150$\,$pc of the Sun using their kinematics. Ages of these groups of dozens of stars range from $\sim$1-800$\,$Myr. The Hyades moving group cluster is generally held to be one of the oldest known moving groups, with an estimated age between 600 and 800$\,$Myr \citep{1998A&A...331...81P}, while the age of the HR1614 is estimated to be $\sim$2$\,$Gyr \citet{2007AJ....133..694D}. 
With {\it Gaia} data release 2 (\citet[Gaia DR2,][]{2018arXiv180409365G}) including precise proper motions, parallaxes and radial velocities, the quest to find all nearby ($\lesssim$160$\,$pc) young moving group members down to substellar types has become feasible \citep[e.g.][]{2018arXiv180511715G}. 

However, most moving groups survive up to a few hundred million years before they gradually dissolve when passing close to nearby stars and giant molecular clouds. Consequently, assuming that multiple star-forming events took place in the Solar neighbourhood not so long ago, there might be numerous field stars showing signs of youth in their spectra. With the arrival of various and complementary big spectroscopic stellar surveys such as {\it Gaia} RVS \citep{2018A&A...616A...6S}, APOGEE \citep{2017AJ....154...94M}, SEGUE~I and II \citep{2009AJ....137.4377Y}, SDSS-V \citep{2017arXiv171103234K}, RAVE \citep{2017AJ....153...75K}, LAMOST \citep{2015RAA....15.1095L}, Gaia-ESO \citep{2012Msngr.147...25G}, and the future FunnelWeb survey (Rains et al. in prep.), 4MOST \citep{2012SPIE.8446E..0TD}, WEAVE \citep{2016ASSP...42..205D} and MOONS \citep{2016ASPC..507..109C}, systematic searches for young field stars on large scales became possible.
For example, \citet{2017ApJ...835...61Z} found $\sim$2000 field stars from the RAVE Survey \citep{2017AJ....153...75K} to be younger than 100$\,$Myr based on their Ca~II infrared triplet excess emission.


Other spectral features of low-mass young stars include H$\alpha$ emission \citep{2008AJ....135..785W,2015AJ....149..158S} that remains present up to a few billion years \citep{2008ApJ...687.1264M} and chromospheric emission in Ca~II~H\&K \citep[e.g.][]{1984ApJ...279..763N,2008ApJ...687.1264M,2015RAA....15.1282Z}. 
For a review see \citet{2010ARA&A..48..581S}.

To constrain ages of the youngest, often pre-main sequence (PMS) objects in the first few tens of millions of years of their lives, the presence of the Lithium 6707.78{$\,$\AA} absorption line is used.
As PMS objects contract toward the zero-age main sequence and their core temperature reaches $\sim$3$\,$MK, $^7$Li ignites on a mass-dependent timescale. Because the mixing timescale in fully convective PMS stars is short, Lithium becomes depleted throughout the star almost immediately ($<$5\% of stellar age; 
a few to ten Myr, \citealt[][]{2014EAS....65..289J}). Ignition depends on stellar mass and creates a narrow boundary (Lithium depletion boundary) in luminosity between Lithium depleted stars and those with lower luminosities that still retain their initial Lithium abundance \citep{2014prpl.conf..219S}. While the technique can be used for groups of stars with the same age, only upper age limits can be estimated for field stars between 20-200$\,$Myr. 

This work is focused on the GALAH\footnote{https://galah-survey.org/} data (The Galactic Archaeology with HERMES survey, \citep{2015MNRAS.449.2604D,2017MNRAS.465.3203M,2018MNRAS.478.4513B}, an ongoing large-scale stellar spectroscopic survey of the southern sky that currently contains over 800,000 spectra \citep{2015MNRAS.449.2604D}, among which 98,000 are from the K2 and 45,000 from the TESS programes with HERMES. 
The Lithium line in the red channel (6478--6737{$\,$\AA}) is used to search for young nearby field K and M dwarfs. 

The GALAH dataset is presented in Section \ref{sec.data}. Data analysis and determination of equivalent widths of Lithium absorption line is described in Section \ref{sec.ew}. Lithium-strong pre-main sequence stars and the identification of young stars above the main sequence using broad-band photometry and {\it Gaia} parallaxes alone, utilizing Lithium as a test of the procedure, are discussed in Section \ref{sec.discussion}. Conclusions are given in Section \ref{sec.conclusions}.

\begin{figure}
\includegraphics[width=\linewidth]{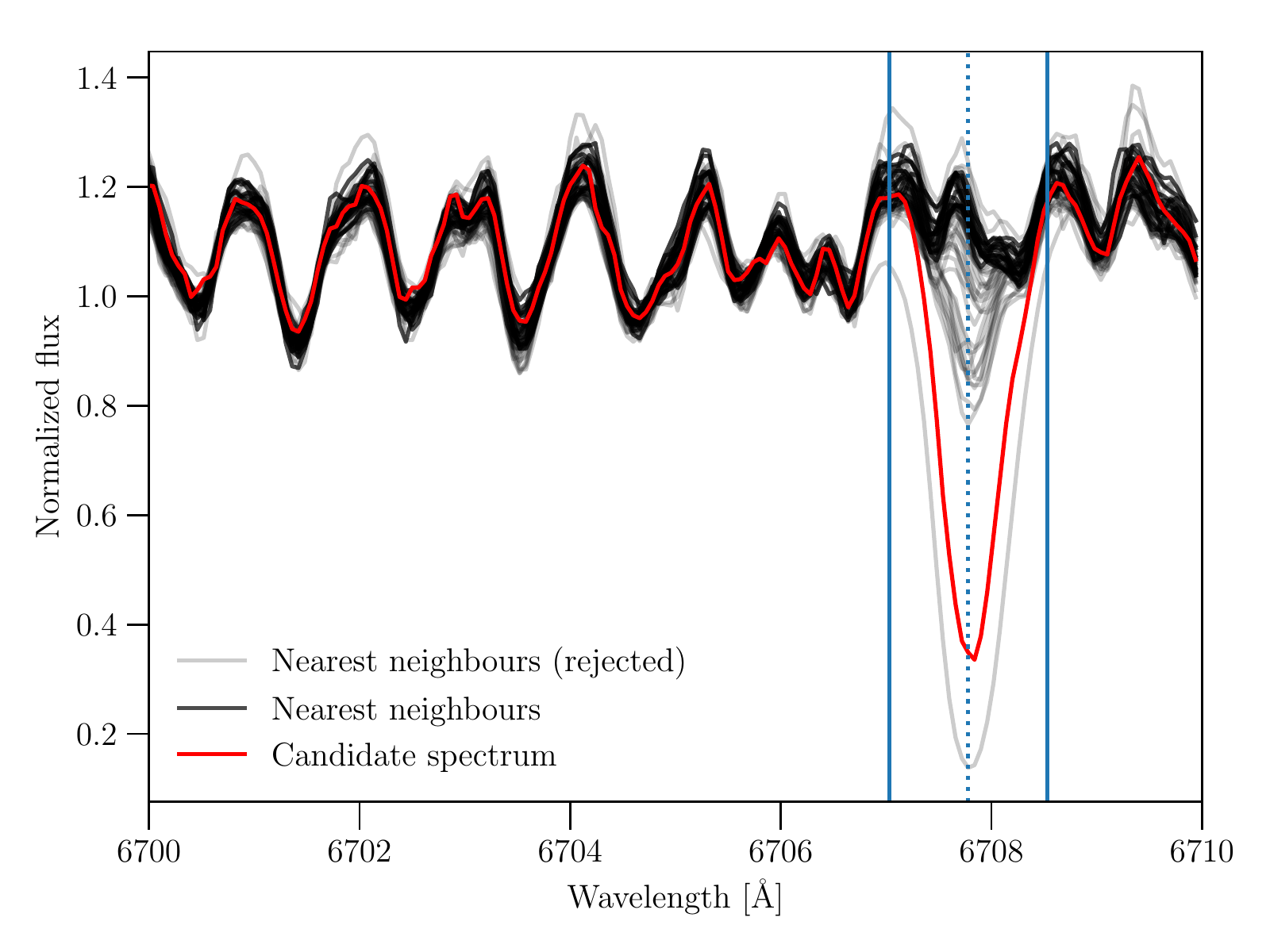}
\caption{An example of a candidate spectrum of an early M dwarf and its nearest neighbours. All spectra are normalized using the same algorithms for all stellar types. Only neighbours with the highest central flux values (at 6707.78{$\,$\AA}) were kept, the rest were rejected to remove objects with strong Lithium. Equivalent width is determined within solid blue lines while dashed line is centered at the Li line.} 
\label{fig.knn_example}
\end{figure}

\section{Data} \label{sec.data}
GALAH is an ongoing large-scale spectroscopic survey that makes use of the multi-object ($\sim$392) spectrograph HERMES \citep{2010SPIE.7735E..09B} at the 3.9-metre Anglo-Australian Telescope, located at the Siding Spring Observatory, Australia. The high-resolution spectrograph ($R\sim$28,000) is composed of four optical channels covering 4713--4903{$\,$\AA}, 5648--5873{$\,$\AA}, 6478--6737{$\,$\AA}, and 7585--7887{$\,$\AA}. All spectra are continuum normalized. The internal database exceeds 800,000 spectra of 600,000 stars collected between January 2014 and February 2018. The survey is magnitude limited, with most of the stars between $V_{JK}$=12 and 14. 
Additionally, crowded galactic plane was avoided (|b|$>10^{\circ}$), and $-80^{\circ}<\delta<+10^{\circ}$. 
The median value of signal-to-noise per resolution element (S/N) in the red arm for candidate spectra is 80, with 90\% of stars having S/N better than 40. For more details, including stellar parameter and abundance determination in GALAH, see B18.

\begin{figure*}
\includegraphics[width=\linewidth]{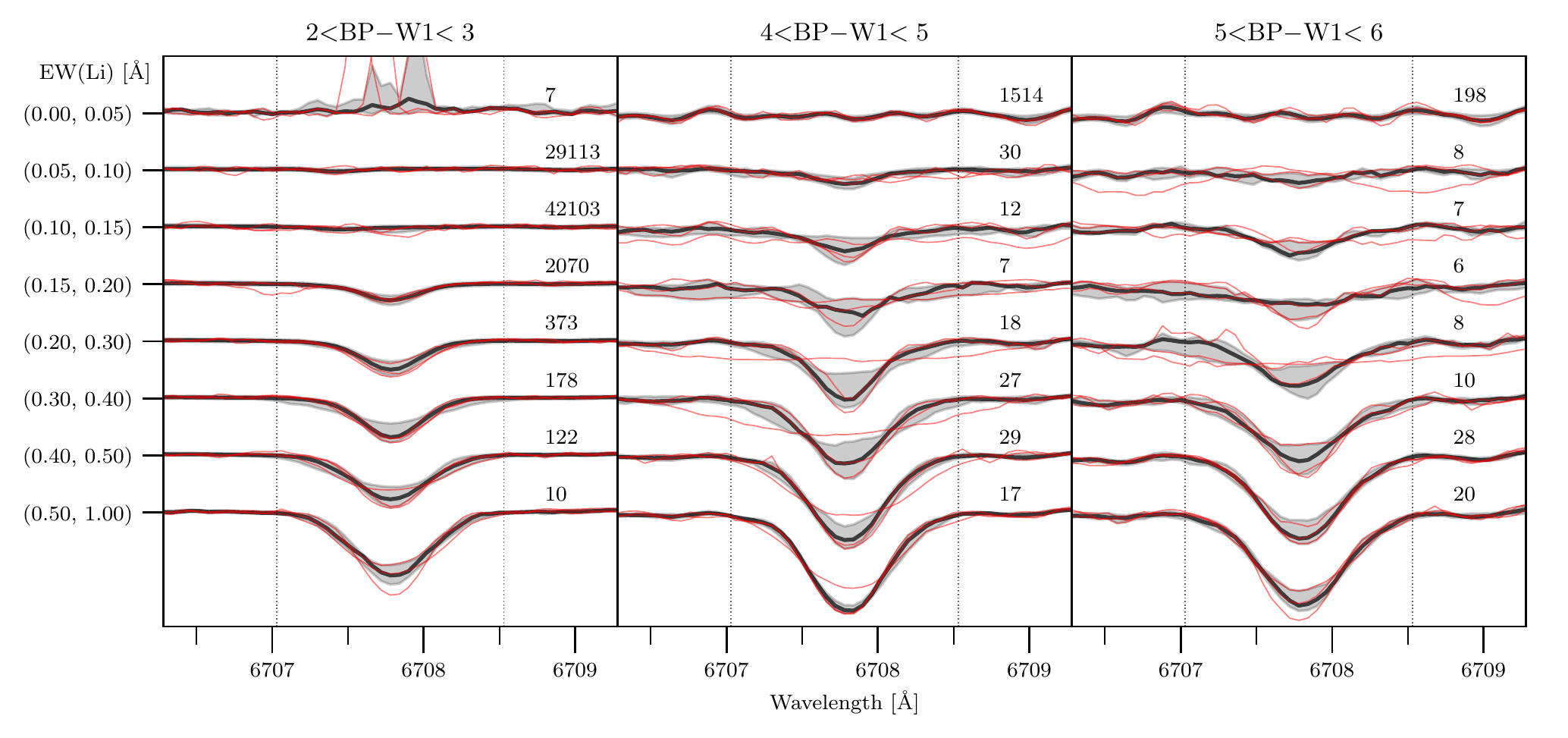}
\caption{Median spectra for BP-W1 and EW(Li) bins with $1\sigma$ deviations. Random individual spectra from the corresponding bins are overplotted in red. Spectra with S/N$<$40 are excluded.
Number of spectra averaged in each bin is given on the right hand side above each median spectrum.}
\label{fig.average_spectra}
\end{figure*}

Here we study red arm spectra (including the Lithium $\lambda_{Li}=6707.78${$\,$\AA} line) of 438,888 
dwarfs to determine the presence of Lithium absorption line. The sample was chosen from 800,000 spectra by exclusion of stars with $\log{g}<3.75$ and $T_{\mathrm{eff}}>7000\,\mathrm{K}$. As not all stars have determined atmospheric parameters, absolute {\it Gaia} G magnitude (G$<$2) and BP-RP$<$0.5 were used to additionally vet the data. The rest of the stars from the internal database with available radial velocities were kept in the candidate list.

Additionally, manually selected spectra classified as peculiars in GALAH were attached to the list.
The list of peculiar spectra in GALAH -- spectra with features like emission lines, broad TiO bands etc., including spectra affected by technical issues -- has been prepared with the t-SNE classification technique \citep{2017ApJS..228...24T}. As the pipelines being used to analyze GALAH data are in some cases not able to treat such spectra, objects from the peculiar list have been excluded from further consideration in the main GALAH survey. However, 5132 very cool stars in this work (mostly M dwarfs) have been selected manually from the list of peculiars with the help of the t-SNE projection and added to the list of Lithium-strong candidates. The vetting has been done by the eye inspection as missing radial velocities for these spectra made the t-SNE projection less efficient.
Binary stars have not been treated separately. The fraction of spectra with other types of peculiarity in the sample used in this paper is about 5$\,$\%.

Radial velocities for the manually selected subsample of 3519 cool peculiars have been determined with cross-correlation using the same pipeline as for the main GALAH survey. The synthetic library AMBRE \citep{2012A&A...544A.126D} has been expanded with inclusion of additional cool templates (3500, 3700 and 3900$\,$K). Metallicity and surface gravity were fixed for all templates (0 and 4.5$\,$dex, respectively). For more details see \citet{2017MNRAS.464.1259K}.
In total, 434,215 spectra of 400,302 unique objects are used in this work. The sample consists of three different subsets: 350,901 GALAH, 62,551 K2 and 28,955 TESS spectra from the HERMES programs.

\section{Lithium-strong dwarfs} \label{sec.ew}
The main goal of the GALAH survey is to address the chemical abundances of the stars in the Milky Way galaxy. Lithium abundance has been determined for the majority of the stars in the GALAH database. However, due to difficulties in modeling M dwarfs with strong TiO bands and a subsequent lack of a reliable training set, Lithium data is unavailable below $\sim$4000$\,$K. This work presents an independent estimate of the equivalent widths (EW) of the Lithium line for low-mass stars. The same method is used for both K and M dwarfs to provide consistent measurements across the entire parameter space. A data-driven and model-free approach to obtain equivalent widths from the residual spectra (template minus candidate) is exploited to produce robust and reliable values.

\subsection{Templates}
To find a corresponding best-matching synthetic template, the atmospheric parameters of the candidate star needs to be known. The estimation of atmospheric parameters in cool stars is complicated and an alternative model free approach is needed. Because GALAH is a large survey with nearly one million spectra (with about two thirds being spectra of dwarfs), it is straight-forward to take advantage of data-driven methods to search for templates for all types of stars from the hottest to the coolest parts of the Hertzsprung-Russell diagram.

In order to find best-matching template spectra for every candidate star, a nearest neighbour search\footnote{Utilizing KDTree.} among normalized fluxes was performed in a pool of 434,215 candidate spectra. Neighbours were searched among candidate spectra themselves.

All GALAH spectra are shifted to the rest frame and interpolated to match the same equidistant wavelength bins ($\Delta \lambda$=0.06{$\,$\AA}). All of them are continuum normalized using the same procedure, and the assumption is that similar stars have similar continuum values, including the depth of the molecular bands of the coolest stars. Despite the fact that four different spectral bands are available in GALAH, only the red band (covering the Lithium line) was used in the nearest neighbour search. Only selected parts of the spectrum that carry most of the information (27 regions with the strongest lines of the early K spectrum in the red HERMES arm, Fig. \ref{fig.bands}) were included in the search. 
Lithium and H$\,\alpha$ lines were excluded to enable the search for stars with identical spectra but no signs of youth in these two regions.

One hundred nearest neighbours have been determined for each candidate star.
The differences between the colors of candidate spectra and their nearest neighbours are rather small. For the first 50 nearest neighbours, the majority of stars are within (-0.32, +0.14; corresponding to approximately -100, +50$\,$K) from the star in BP-W1 colour.

\subsection{Equivalent widths}
The equivalent width of each Lithium line is determined from the residual spectra. Each candidate spectrum is subtracted from each of its nearest neighbours to determine residual spectra.
The procedure has two iterations. The steps of the first iteration are:
\begin{enumerate}
\item Find suitable neighbours. \label{s1}
\item Correct continuum slope of the candidate spectrum to match its neighbours. \label{s2}
\item Subtract candidate spectrum from its neighbours to get residual spectra. \label{s3}
\item Shift residual spectra in flux so that their median outside Lithium line is zero. \label{s4}
\item Find equivalent widths for all residual spectra. \label{s5}
\item Determine EW(Li) and uncertainties. \label{s6}
\end{enumerate}

\begin{figure*}
\includegraphics[width=\linewidth]{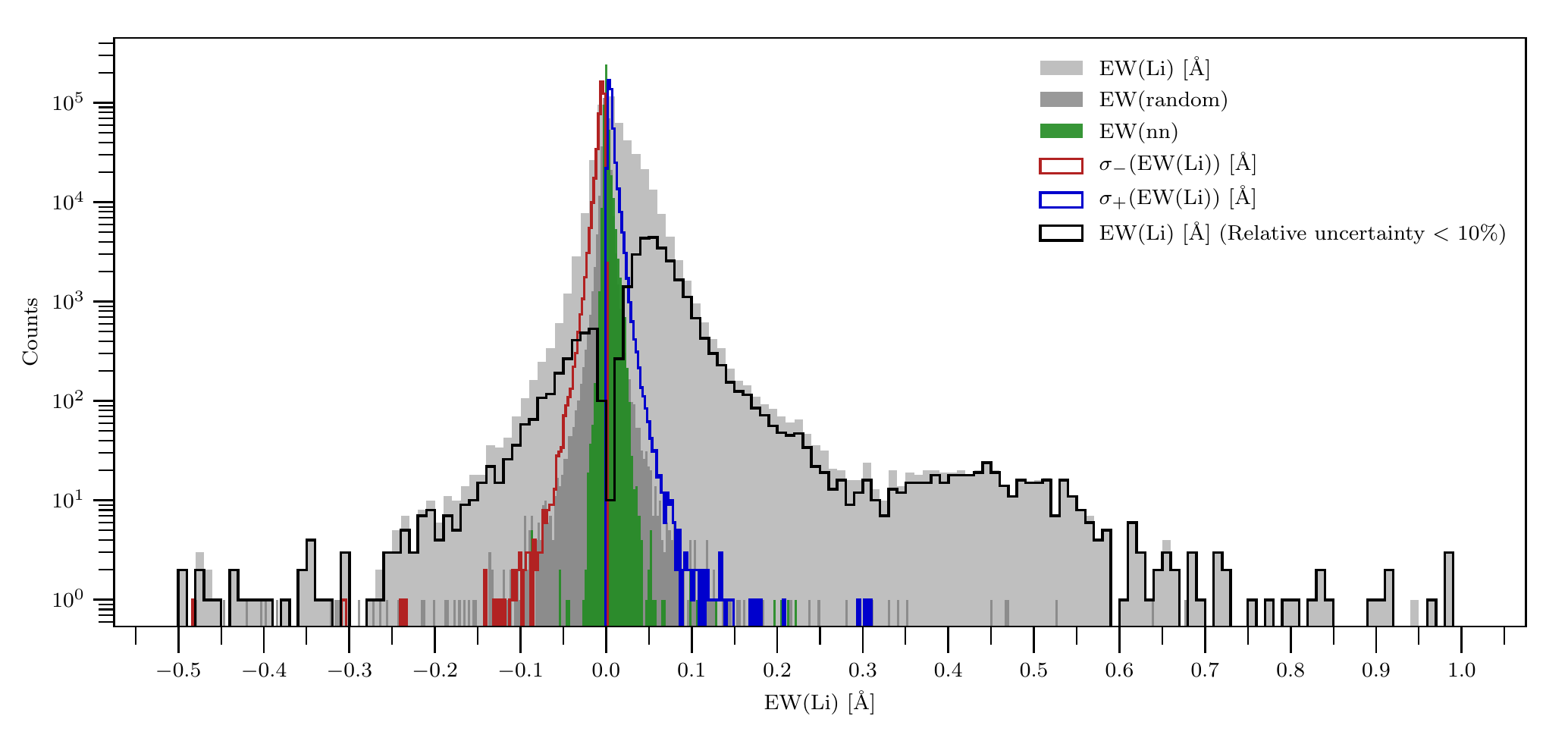}
\caption{Distribution of equivalent width of Lithium EW(Li). Most of the stars are Lithium-weak, but the distribution is bimodal. EW(random) is a measure of fundamental noise and template mismatch uncertainty ($\pm$0.003{$\,$\AA}). EW(nn) corresponds to Lithium-weak templates (nearest neighbours).
Note that the y-scale is logarithmic.}
\label{fig.ewli_distribution}
\end{figure*}

In step \ref{s1}, a candidate's first 100 nearest neighbours are investigated. Those with signal-to-noise ratio less than 40 are rejected. Then, in order to find Lithium-weak stars, neighbours are ordered with respect to their flux in the center of the Lithium line. Only those between the 66th and 95th percentile are kept in the pool. A typical number of accepted neighbours is 35 (with an imposed hard upper limit of 50) and they are assumed to represent continuum within the Lithium range. From this point on, only wavelengths between 6690 and 6710{$\,$\AA} around the Lithium line itself are taken into consideration. The procedure is illustrated in Fig.~(\ref{fig.knn_example}).

Continuum normalization, especially in the case of very cool stars with wavy pseudo-continuum levels, is sometimes not sufficient at the edges. For this reason, the slope of the candidate is corrected to make sure it overlaps with its nearest neighbours in step \ref{s2}. Slope mismatch occurs only at the edge and does not affect the nearest neighbour search significantly.

The candidate spectrum is subtracted from each nearest neighbour in step \ref{s3}. Residual spectra are shifted along the y-axis to match a zero median flux value outside the Lithium line (step \ref{s4}). Equivalent widths are determined for each ($n$-th) residual spectrum (step \ref{s5}) as
\begin{equation}
\mathrm{EW(Li)}_n = \sum_{i=0}^{N} r_{i} \Delta \lambda
\end{equation}
where $r_i$ is the residual flux at the i--th pixel and N=25 is number of pixels in the Lithium line between $\lambda_{Li}$-0.75{$\,$\AA} and $\lambda_{Li}$+0.75{$\,$\AA}.
Finally, the EW(Li) of the candidate spectrum is given as a median value of the equivalent widths EW(Li)$_n$ (step \ref{s6}). 

Uncertainties are reduced with the second iteration. The latter differs from the iteration 1 only in the step \ref{s1} where selection of the most Lithium-weak neighbours is based on their EW(Li). In particular, only $\sim$ 15 neighbours with EW(Li) closest to 0 are accepted for every candidate spectrum. The typical EW(Li) of selected nearest neighbours is 8$\cdot 10^{-5}${$\,$\AA}.

The typical uncertainties of EW(Li) are $\pm$0.005{$\,$\AA} and $\pm$0.014{$\,$\AA} for M dwarfs. This is only slightly higher than the uncertainties originating from template mismatch and noise ($\pm$0.003{$\,$\AA}), determined from 10 different random intervals outside Lithium (but with same width as Lithium range). Additional uncertainty in Lithium might originate from blending of the CN and Iron~I line (6707.441{$\,$\AA}). 
Figure (\ref{fig.average_spectra}) shows an increasing strength of the Lithium line in median spectra within EW(Li) bins for four different colour ranges. Although typical EW(Li) uncertainties are small, the Lithium line starts to stand out around 0.1{$\,$\AA} (Fig. \ref{fig.average_spectra}) for stars with BP-W1$>$4. 
The distribution of EW(Li) (Fig. \ref{fig.ewli_distribution}), centered at 0.007{$\,$\AA}, is an order of magnitude broader than uncertainties (-0.01, +0.03){$\,$\AA}.

\begin{figure*}
\includegraphics[width=\linewidth]{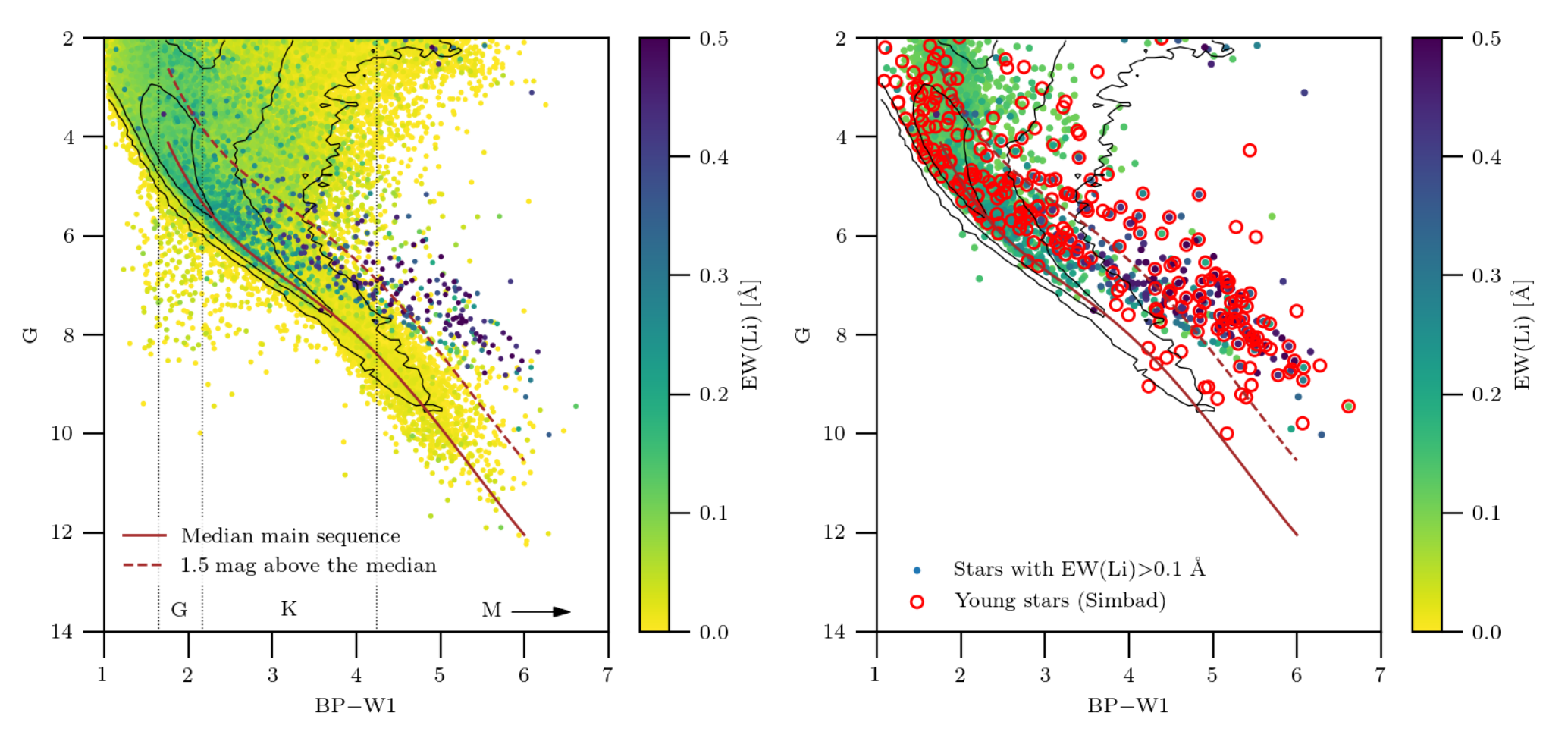}
\caption{\textit{Left:} Colour-magnitude diagram with absolute G magnitude versus BP-W1 for stars with relative parallax errors better than 20$\,$\% and EW(Li)>0.037{$\,$\AA}. The majority of Lithium-strong stars lie above the main sequence. Number density of stars is designated by the black contours (10, 100 and 1000 stars per 0.1~$\times$~0.1 magnitude bins). Spectral types are marked with vertical lines; BP-W1$\approx$6 corresponds to M4 stars.
\textit{Right:} Stars with EW(Li)>0.1{$\,$\AA} and stars with young flags in Simbad (red circles; pre-main sequence, X-ray, T~Tau, young stellar object). Simbad young stars with missing EW(Li) values in the plot have either EW(Li)$<$0.1{$\,$\AA} or relatively large parallax error.}
\label{fig.cmd}
\end{figure*}

\begin{figure*}
\includegraphics[width=\linewidth]{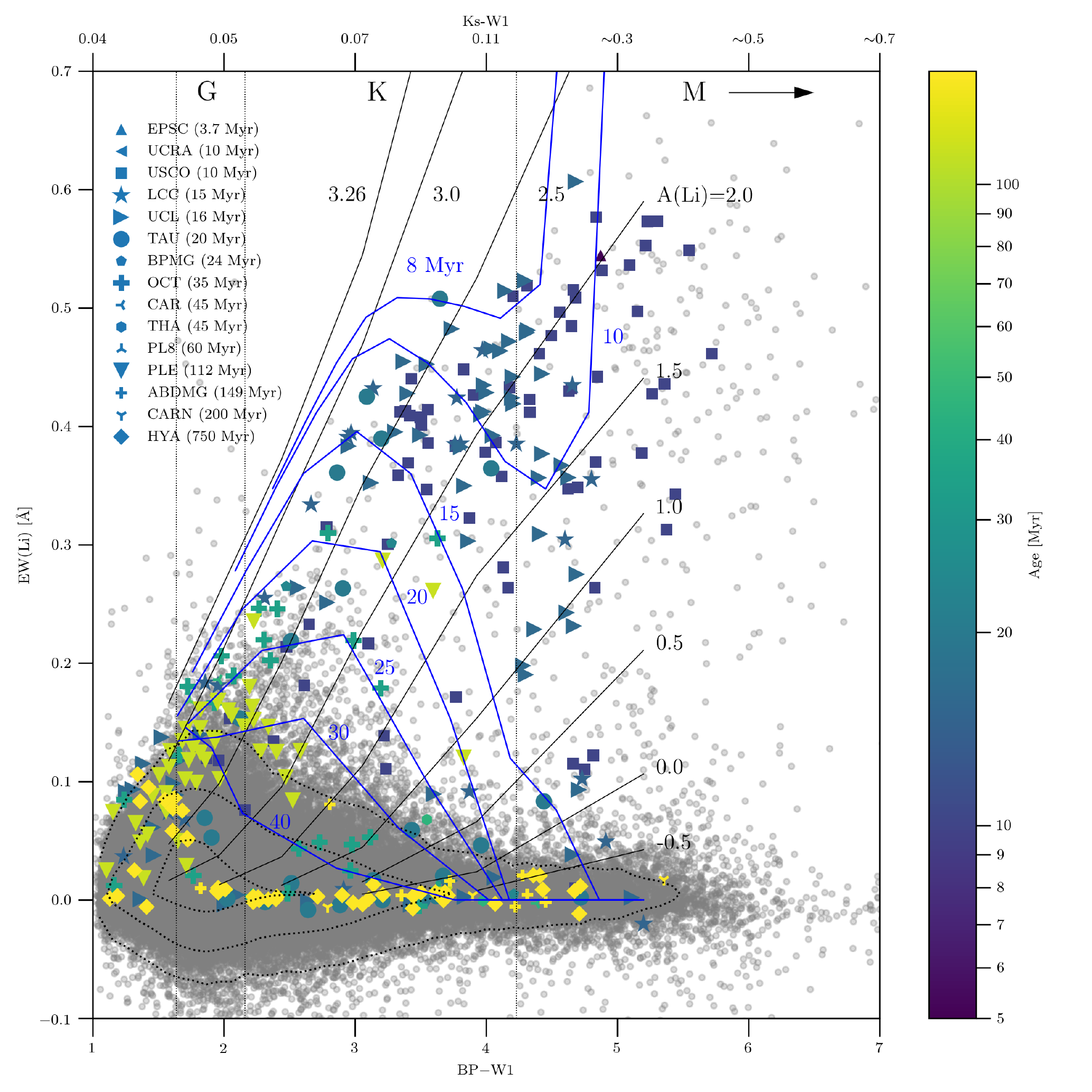}
\caption{The dependence of the measured EW(Li) on the colour. Contours designate number density (100, 1000 and 10000 stars per 0.1$\,$mag~$\times$~0.04{$\,$\AA}). Curves of growth for selected absolute abundances A(Li) are shown with solid black lines. Isochrones tracking Lithium depletion are shown in blue. GALAH bonafide and candidate members of the known moving groups are colour-coded with their age.} 
\label{fig.colourmg}
\end{figure*}

\section{Discussion} \label{sec.discussion}
The distribution of EW(Li) is shown in Figure (\ref{fig.ewli_distribution}). The majority of stars show no Lithium absorption line, as expected. The center of the distribution is at $0.007_{-0.012}^{+0.028}${$\,$\AA}. However, the distribution does not only seem asymmetrical, with an overabundance of Lithium-strong stars, but is also bimodal with the second peak centered at EW(Li)$\sim$0.45{$\,$\AA} and with a width of approximately $\pm$0.2{$\,$\AA}. A few stars are found in the higher range from 0.7 to 1{$\,$\AA}. 
If the distribution was unimodal and normal with the center at 0.007{$\,$\AA} and $\sigma=0.012 \,${\AA} ($\sigma$ measured for the left (negative) side of the distribution), then the number of stars should have fallen below 1 at EW(Li)=0.1{$\,$\AA}. While the assumption of normal distribution is not entirely true (the number of stars with negative EW(Li) falls to 0 around -0.3{$\,$\AA}), there are 3389 stars with EW(Li)$>$0.1{$\,$\AA}, 760 above 0.2{$\,$\AA}, 376 above 0.3{$\,$\AA} and 102 with EW(Li)$>$0.5{$\,$\AA}. 

The colour-magnitude diagram (Fig. \ref{fig.cmd}), using {\it Gaia}~DR2 parallaxes \citep{2016A&A...595A...1G} reveals the nature of stars from the higher EW(Li) peak. Most such stars are M dwarfs with BP-W1$>\sim$4 which reside above the main sequence. Their EW(Li) is an order of magnitude higher than main sequence stars of the same colour. As stars with the same temperature would have similar levels of blending and pseudo-continuum depression, it can be concluded that these effects are negligible in this work. Equivalent width is thus a robust measure for the strength of the Lithium absorption line for the purpose of discovering young stars. 
Table \ref{tab.ewli} lists measured EW(Li) and uncertainties for all 434,215 stars from this work.

\begin{table}
\begin{tabular}{cccc}
Designation & EW(Li) & $\sigma$(EW(Li))$_-$ & $\sigma$(EW(Li))$_+$ \\
{\textit Gaia} DR2 & $\mathrm{\mathring{A}}$ & $\mathrm{\mathring{A}}$ & $\mathrm{\mathring{A}}$ \\
 \hline
6209522875989260032 & 0.021 & 0.002 & 0.007 \\
5915739214516226688 & 0.056 & 0.005 & 0.006 \\
6209923617916698240 & 0.038 & 0.008 & 0.005 \\
6172208513655971072 & 0.033 & 0.002 & 0.005 \\
6170704519188425600 & 0.012 & 0.007 & 0.010 \\
\end{tabular}
\caption{List of GALAH stars with measured Lithium equivalent width. Full version with 434,215 spectra is available online.}
\label{tab.ewli}
\end{table}

For a reference, \texttt{astroquery} \citep{astroquery} was used to cross-match GALAH with the Simbad catalogue using 2MASS identifiers. Objects marked as young (e.g. 'T Tau' - T Tau-like object, 'pr' - pre-main sequence star, 'Y*O' - young stellar object, and candidates) overlap with the Lithium-strong sequence in Fig. (\ref{fig.cmd}, right).

\vspace{1em}
At a given abundance, the line strength increases with decreasing effective temperature as shown in Fig.~\ref{fig.colourmg}. This is mainly due to the population of the ground state increasing thanks to a decreasing ionization fraction.
We plot there indicative non-LTE equivalent widths from \citet{1996A&A...311..961P}, computed for solar metallicity and shown here for $\log g = 4.5$ over a broad range of temperatures and Lithium abundances. While these calculations use outdated model atmospheres and neglect transitions due to collisions with neutral hydrogen atoms, more recent work on FGK type stars taking into account accurate transition rates \citep[e.g.][]{2009A&A...503..541L} generally predicts larger equivalent widths, with differences at most 20\,\% for the relevant parameter space. Additionally, variations in surface gravity affect line strengths (and abundances) by as much as 10\,\%. 

We also show indicative spectral types in the figure. These are based on the temperature-spectral type relation given by \citet{2013ApJS..208....9P}\footnote{In the version from 2018.08.02, available online: \url{http://www.pas.rochester.edu/~emamajek/EEM_dwarf_UBVIJHK_colors_Teff.txt}}. As empirical compilations of BP--W1 are not yet available, we use synthetic spectra to compute colours over a range of effective temperatures, and shift the zero-point to reproduce the colors of the Sun. Comparing our synthetic colours to the tabulated values of B--W1 from \citet{2013ApJS..208....9P}, we find agreement to within 0.1 mag for FGK types, with increasing errors as large as 0.4 mag for mid-M types.

\vspace{1em}
The plot reveals that for M- and late K dwarfs any amount of Lithium left in the atmosphere will still show up at EW(Li)$>$0.1{$\,$\AA} while only high abundances can be easily detected in G dwarfs.
The change of abundance in late K and early M stars affects EW(Li) to a much greater extent than in early K dwarfs. In addition to this, the depletion rate depends on stellar mass and occurs on shorter timescales for early M than K dwarfs \citep{2008ApJ...689.1127M,2014prpl.conf..219S}. 
Assuming the initial absolute abundance of 3.26 \citep{2009ARA&A..47..481A}, the \citet{2015A&A...577A..42B} models of Lithium depletion combined with the \citet{1996A&A...311..961P} curves of growth were used to derive Lithium isochrones. 

All M and late K dwarfs with detectable Lithium are younger than 15-20$\,$Myr (30$\,$Myr, respectively) and thus still in their pre-main sequence phases.

GALAH known and new candidate moving group members (378 stars) found by the Banyan~$\Sigma$ code (G18) 
have been color-coded with their age (their membership is discussed in the next subsection).
Similar stars with the same nominal age are rather scattered in the plot, well beyond the internal measurement error of this work. According to \citet{2010ARA&A..48..581S}, much of the scatter for the coolest stars is real, but its origin is not yet confirmed. Recently, \citet{2018A&A...613A..63B} have shown that low-mass fast rotators in the Pleiades are systematically rich in Lithium compared to slowly rotating members of the cluster. 
Additionally, binary stars have not been treated separately in this work, and spectra of some stars (especially pre-main sequence, e.g. T~Tauri objects) might be affected by veiling that causes lines to appear shallower.

\begin{table}
\begin{tabular}{lccc}
Name & Bonafide & New candidates & EW(Li)$>$expected \\
\hline
USCO  & 14 & 65 & 55 \\
UCL  & 2 & 65 & 51 \\
LCC  & 3 & 25 & 16 \\
PLE  & 31 & 36 & / \\
HYA  & 34 & 25 & / \\
OCT  & 0 & 41 & 17 \\
TAU  & 4 & 21 & 4 \\
ABDMG  & 0 & 8 & / \\
IC2602  & 1 & 6 & 0 \\
BPMG  & 0 & 3 & 3 \\
CARN  & 0 & 3 & / \\
CAR  & 0 & 2 & 1 \\
THA  & 0 & 2 & 1 \\
TWA  & 0 & 1 & 0 \\
PL8  & 0 & 1 & 1 \\
UCRA  & 0 & 1 & 1 \\
EPSC  & 1 & 0 & 0 \\
ROPH  &1  & 0 & 0 \\
\hline
Total & 91 & 305 & 150
\end{tabular}
\caption{Summary of bonafide (G18) and new Banyan~$\Sigma$ candidate members in GALAH. Their EW(Li) was compared with the expected values for their age; 'EW(Li)$>$expected' lists number of stars with EW(Li) equal or higher than expected values. No info is provided for groups above 100$\,$Myr or those with stars outside the isochrone interpolation colour range.} 
\label{tab.new_members}
\end{table}

\begin{figure}
\includegraphics[width=\linewidth]{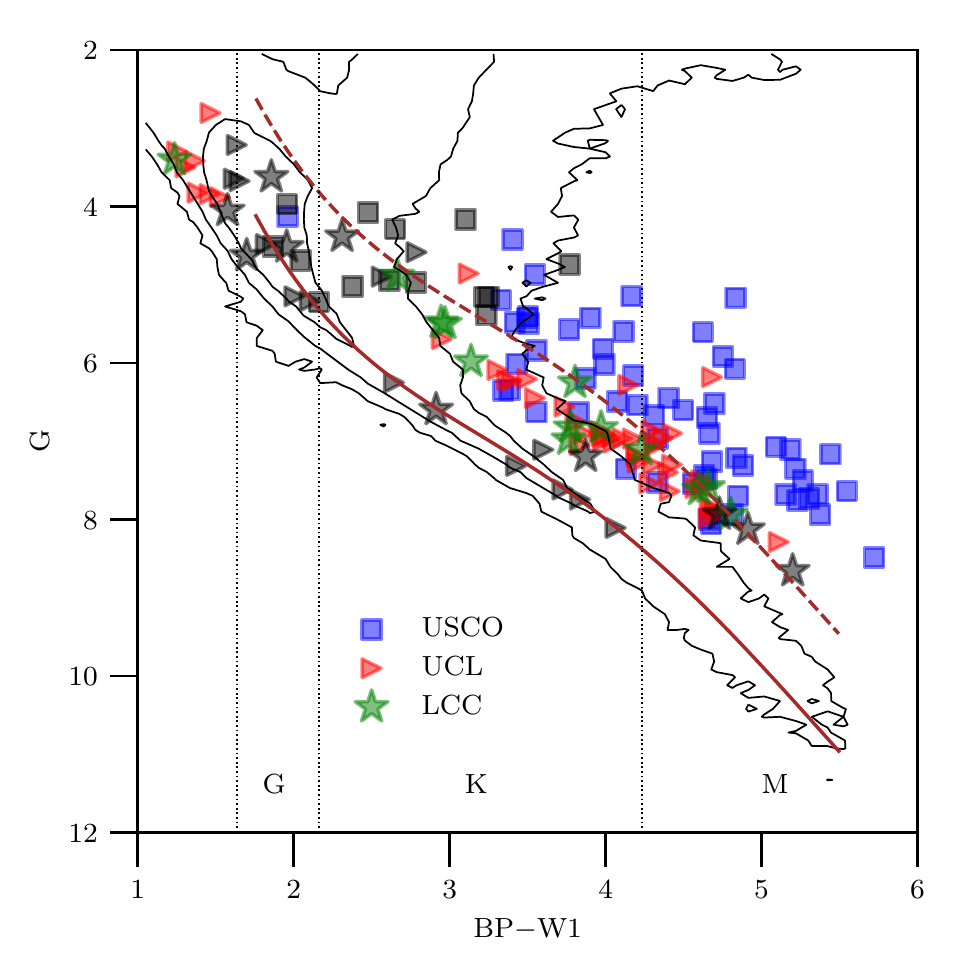}
\caption{Bonafide and new candidate members of the Scorpius-Centaurus association with measurable Lithium line. Candidates with questionable membership due to their EW(Li) (and in some cases their evolutionary stage) not corresponding to the expected values for their colour and age are shown in black.}
\label{fig.scocen}
\end{figure}

\subsection{New candidate members of young associations}
The crossmatch between the G18 bonafide members and the entire GALAH sample using stellar designations revealed 91 stars that are known members of the young associations. However, since GALAH provides radial velocities \citep{2018MNRAS.481..645Z,2018MNRAS.478.4513B} and completes the 6D positional and kinematic information, 305 new candidate members have been found with the Banyan~$\Sigma$ code (G18) using parallaxes and proper motions from {\it Gaia}. Only stars with radial velocity uncertainties smaller than 1$\,$km$\,$s$^{-1}$ were considered. Note that M dwarfs from the manually selected subsample of 3519 spectra (Sec. \ref{sec.data}) have not been included in the Banyan~$\Sigma$ membership test due to their relatively large uncertainties in radial velocities. 

Membership probabilities are always more than 0.5; in fact, most of them are practically 1. 
However, it is clear that for individual stars, a high membership probability of an association from Banyan~$\Sigma$ does not in fact mean that the star is likely to be young. For example, only 19$\,$\% candidates of the Taurus-Auriga (TAU; 1-2$\,$Myr in G18, but \citet{2017ApJ...838..150K} argue that it is host to a distributed older population with 10-20$\,$Myr; 20$\,$Myr isochrone is used in this work) and 41$\,$\% Octans (OCT) candidates show Lithium strength equal or greater than expected at their age (Fig. \ref{fig.colourmg}). 
In the entire sample of candidates, only 150 reside above their expected isochrone.
Because a more detailed analysis of the membership is beyond the scope of this paper, we merely summarize results from the Banyan~$\Sigma$ code (Table \ref{tab.new_members}) and provide a ful candidate list (Table \ref{tab.associations}).

Half of the new candidates (154) are related to the Scorpius-Centaurus association. Figure \ref{fig.scocen} shows colour-magnitude diagram of the candidate stars for each of the three subgroups (Upper Scorpius - USCO, Upper Centaurus-Lupus - UCL and Lower Centaurus-Crux - LCC). Although most of them show EW(Li) above the detection threshold, the inconsistency with the expected Lithium depletion at their age in some cases makes their membership questionable. 123 stars though have EW(Li) above the expected values.

\subsection{Young stars from absolute magnitudes}\label{sec.absolutemag}
High EW(Li) is found in young cool stars. At the same time, such M stars reside above the main sequence. This section addresses the possibility of identifying young stars from the broad-band photometry and the {\it Gaia} parallaxes alone using Lithium as a test of the procedure.
To better understand the relation between Lithium strength and the location of young stars on the colour-magnitude diagram, the main sequence itself must first be parameterized. 
A polynomial of the 5th order was fitted to the median values of the colour in each of the bins in the absolute G magnitude (binsize 0.2$\,$mag in G) for $G>4$ and BP-W1$<6$. Effectively, this polynomial is a curve connecting the densest parts of the main sequence. The fit was performed for all filter combinations among (BP, G, RP, J, H, K$_s$, W1, W2). BP, G and RP are obtained from the {\it Gaia} DR2 photometric catalogue \citep{2018arXiv180409368E}, J, H and K$_s$ are 2MASS magnitudes \citep{2006AJ....131.1163S} while W1 and W2 are from the WISE catalogue \citep{2014yCat.2328....0C}.

The BP-W1 colour turned out to have the least dispersion around the main sequence in the G magnitude: 
\begin{dmath}
G = 4.717\cdot 10^{-3} \; c^5 -0.149 \; c^4 + 1.662 \; c^3 - 8.374 \; c^2 + 20.728 \; c - 14.129
\end{dmath}
where c is BP-W1.

If the criterion for stellar youth is EW(Li)$>$0.1{$\,$\AA}, 
then the probability that a star at a selected height above the main sequence is young, can be estimated.
Ideally, we would like to select all young stars of a certain spectral type and keep contamination level as small as possible.
This is easiest to achieve for M dwarfs with BP-W1$>$5 (Fig. \ref{fig.probability}) as their approach toward the main sequence is the slowest. If selecting stars 1.5$\,$mag or more above the main sequence, all young objects are included, but 50\% of all selected stars are Lithium-weak.
Binary levels increase significantly 
below BP-W1$=$5 while a fraction of young stars is still recovered. In other words, selection of young stars above the main sequence is highly effective for M dwarfs, moderately effective for K dwarfs and ineffective for G dwarfs.

Apart from reddening, the spread of the main sequence in luminosity (absolute magnitude) is caused by different metallicities and stellar multiplicity. Binaries are 0.75$\,$mag or less above the main sequence. 
Binaries have not been treated separately, but the presence of Lithium still means that the system is young.

\section{Conclusions} \label{sec.conclusions}
This work addresses the search for nearby young stars using the Lithium 6707.78{$\,$\AA} absorption line.
A simple but robust, model-free approach is used to determine equivalent widths EW(Li) with a data-driven method for G, K and early M dwarfs. The selection of Lithium-weak templates for each candidate spectrum is based on the 27 spectral regions (Fig. \ref{fig.bands} and \ref{fig.knn_example}).
The EW(Li) is given as a median value of equivalent widths within 6707.78$\pm$0.75{$\,$\AA} for each of the (on average) 15 residual spectra. Median spectra for selected colour bins demonstrate increasing strength of Li with EW(Li) (Fig. \ref{fig.average_spectra}).
Typical uncertainties of EW(Li) are $\pm$0.005{$\,$\AA} ($\pm$0.014{$\,$\AA} for M dwarfs) and are comparable with uncertainties originating solely from fundamental noise and template mismatch. Most EW(Li) fall below 0.7{$\,$\AA} with individual cases up to 1{$\,$\AA} (Fig. \ref{fig.ewli_distribution}). 

With this procedure, 3147 stars have been found to have measurable Lithium: 1408 G and 892 K0-K5 (EW(Li)$>$0.1{$\,$\AA}), 335 K5-K9 (EW(Li)$>$0.07{$\,$\AA}) and 512 M0$\sim$M4 dwarfs (EW(Li)$>$0.05{$\,$\AA}) (Tab. \ref{tab.ewli}). Most of the latter reside above the main sequence (Fig. \ref{fig.cmd}). While it is still possible to measure Lithium in M dwarfs with a small fraction of initial Lithium left, a detectable EW(Li) in early K stars traces only Lithium-rich objects (Fig. \ref{fig.colourmg}).

A combination of the Lithium information and 6D stellar kinematics (using GALAH radial velocities) is used to investigate the new candidate membership of the known associations (Tab. \ref{tab.associations}). In particular, we find 305 new candidates, 123 of which belong to the Scorpius-Centaurus association (Fig. \ref{fig.scocen}), 36 are Pleiades cluster and 25 Hyades cluster candidate members.

\begin{figure}
\includegraphics[width=\linewidth]{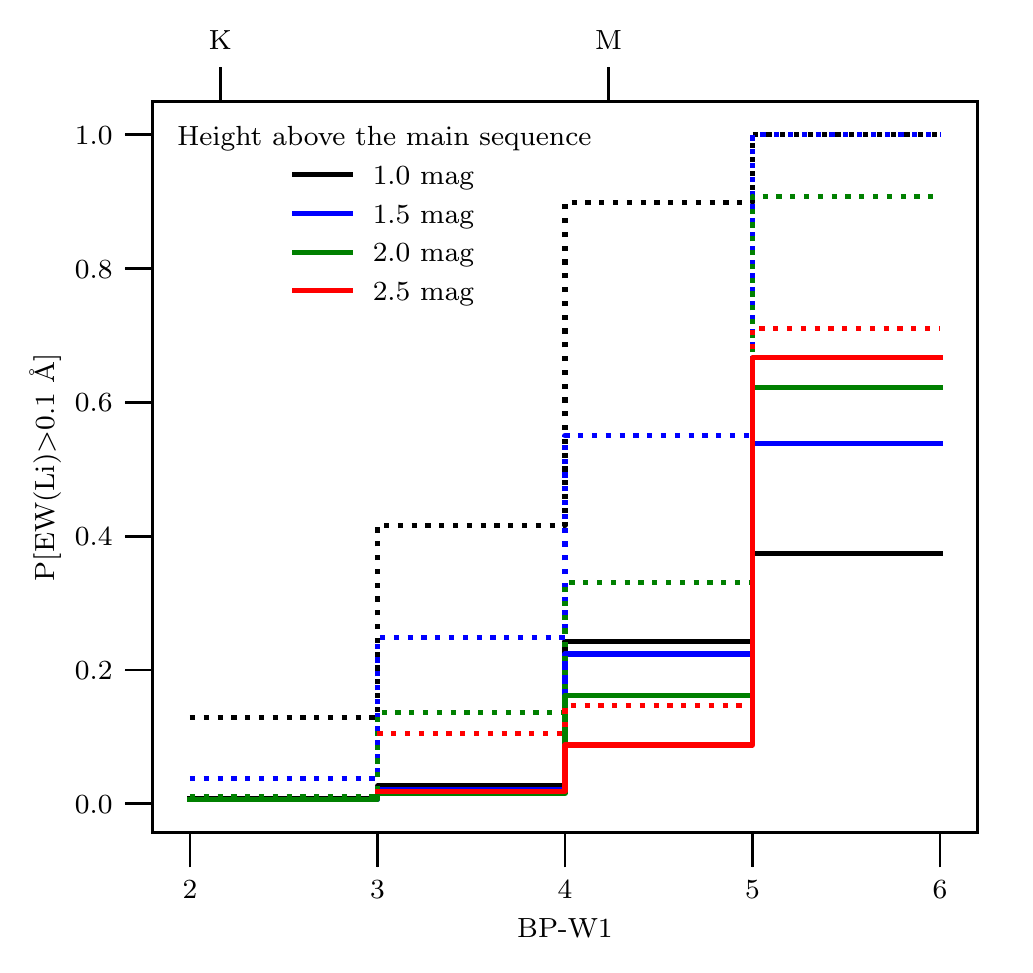}
\caption{Probability that a star is young (EW(Li)$>$0.1{$\,$\AA}) if above the main sequence (solid lines). Dotted lines represent fraction of all young stars in the sample within the colour range residing above a selected height from the main sequence. Selection of young stars above the main sequence is most effective for BP-W1$>$5 as contamination is the lowest and nearly all young stars can be recovered.}
\label{fig.probability}
\end{figure}

Information on the strength of the Lithium line for the majority of the nearby dwarfs in a volume-limited sample is not yet available, but photometric data combined with {\it Gaia} parallaxes to search for objects residing above the main sequence is the most straightforward approach towards a list of all candidate nearby young stars. Figure \ref{fig.probability} investigates contamination levels and completeness for samples of stars with increasing distance from the main sequence. Contamination levels increase significantly (almost 80\%) below BP-W1=5 (M2 dwarfs) while a fraction of young stars are still recovered. To confirm young photometric candidates and search for adolescent stars that have just settled on the ZAMS, large spectroscopic surveys observing Lithium and youth-sensitive regions in all nearby dwarfs are essential in the future (e.g. the FunnelWeb survey (Rains et al., in prep.) and SDSS-V \citealt{2017arXiv171103234K}).

\section*{Acknowledgements}
We acknowledge the traditional owners of the land on which the AAT stands, the Gamilaraay people, and pay our respects to elders past and present.
This work has made use of data from the European Space Agency (ESA) mission {\it Gaia} (\url{https://www.cosmos.esa.int/gaia}), processed by the {\it Gaia} Data Processing and Analysis Consortium (DPAC, \url{https://www.cosmos.esa.int/web/gaia/dpac/consortium}). Funding for the DPAC has been provided by national institutions, in particular the institutions participating in the {\it Gaia} Multilateral Agreement.
M{\v Z} acknowledges funding from the Australian Research Council (grant DP170102233). TN acknowledges funding from the Australian Research Council (grant DP150100250). TZ and K{\v C} acknowledge financial support of the Slovenian Research Agency (research core funding No. P1-0188 and project N1-0040).
This research made use of Astropy, a community-developed core Python package for Astronomy (Astropy Collaboration, 2018).

\begin{landscape}
\begin{table}
\begin{tabular}{rccccccc}
Designation & EW(Li) & $\sigma$(EW(Li))$_-$ & $\sigma$(EW(Li))$_+$ & Bonafide & Best hypothesis & p & EW(Li)$>$expected \\
{\textit Gaia} DR2 & $\mathrm{\mathring{A}}$ & $\mathrm{\mathring{A}}$ & $\mathrm{\mathring{A}}$ &  &  &  &  \\
\hline

6081729701812391168 & 0.420 & 0.007 & 0.011 &  & LCC & 0.9987 & True \\
6145079232473640960 & 0.092 & 0.003 & 0.005 &  & LCC & 0.9530 & False \\
6045116961039667328 & 0.251 & 0.005 & 0.008 &  & UCL & 0.9922 & False \\
3314109916508904064 & 0.086 & 0.001 & 0.002 & Hyades & HYA & 0.9996 & N/A \\
47917816253918720 & 0.021 & 0.008 & 0.018 &  & HYA & 0.9986 & N/A \\
64979732749686016 & 0.152 & 0.002 & 0.003 & Pleiades & PLE & 0.9999 & N/A \\

\end{tabular}
\caption{List of candidate members of known moving groups based on the Banyan~$\Sigma$ \citep{2018ApJ...856...23G} algorithms. Bonafide stars are confirmed members from their paper. Best hypothesis and probability p are the Banyan~$\Sigma$ results. Where applicable, EW(Li)$>$expected gives information on whether Lithium is at least as strong as expected for stellar age.
Full version is available online.}
\label{tab.associations}
\end{table}
\end{landscape}



\bibliographystyle{mnras}
\bibliography{paper}

\begin{thebibliography}{}
\makeatletter
\relax
\def\mn@urlcharsother{\let\do\@makeother \do\$\do\&\do\#\do\^\do\_\do\%\do\~}
\def\mn@doi{\begingroup\mn@urlcharsother \@ifnextchar [ {\mn@doi@}
  {\mn@doi@[]}}
\def\mn@doi@[#1]#2{\def\@tempa{#1}\ifx\@tempa\@empty \href
  {http://dx.doi.org/#2} {doi:#2}\else \href {http://dx.doi.org/#2} {#1}\fi
  \endgroup}
\def\mn@eprint#1#2{\mn@eprint@#1:#2::\@nil}
\def\mn@eprint@arXiv#1{\href {http://arxiv.org/abs/#1} {{\tt arXiv:#1}}}
\def\mn@eprint@dblp#1{\href {http://dblp.uni-trier.de/rec/bibtex/#1.xml}
  {dblp:#1}}
\def\mn@eprint@#1:#2:#3:#4\@nil{\def\@tempa {#1}\def\@tempb {#2}\def\@tempc
  {#3}\ifx \@tempc \@empty \let \@tempc \@tempb \let \@tempb \@tempa \fi \ifx
  \@tempb \@empty \def\@tempb {arXiv}\fi \@ifundefined
  {mn@eprint@\@tempb}{\@tempb:\@tempc}{\expandafter \expandafter \csname
  mn@eprint@\@tempb\endcsname \expandafter{\@tempc}}}

\bibitem[\protect\citeauthoryear{{Asplund}, {Grevesse}, {Sauval}  \&
  {Scott}}{{Asplund} et~al.}{2009}]{2009ARA&A..47..481A}
{Asplund} M.,  {Grevesse} N.,  {Sauval} A.~J.,   {Scott} P.,  2009, \mn@doi
  [Annual Review of Astronomy and Astrophysics]
  {10.1146/annurev.astro.46.060407.145222}, \href
  {https://ui.adsabs.harvard.edu/#abs/2009ARA&A..47..481A} {47, 481}

\bibitem[\protect\citeauthoryear{{Baines} et~al.,}{{Baines}
  et~al.}{2012}]{2012ApJ...761...57B}
{Baines} E.~K.,  et~al., 2012, \mn@doi [\apj] {10.1088/0004-637X/761/1/57},
  \href {https://ui.adsabs.harvard.edu/#abs/2012ApJ...761...57B} {761, 57}

\bibitem[\protect\citeauthoryear{{Baraffe}, {Homeier}, {Allard}  \&
  {Chabrier}}{{Baraffe} et~al.}{2015}]{2015A&A...577A..42B}
{Baraffe} I.,  {Homeier} D.,  {Allard} F.,   {Chabrier} G.,  2015, \mn@doi
  [\aap] {10.1051/0004-6361/201425481}, \href
  {http://adsabs.harvard.edu/abs/2015A%26A...577A..42B} {577, A42}

\bibitem[\protect\citeauthoryear{{Barden} et~al.,}{{Barden}
  et~al.}{2010}]{2010SPIE.7735E..09B}
{Barden} S.~C.,  et~al., 2010, in Ground-based and Airborne Instrumentation for
  Astronomy III. p. 773509, \mn@doi{10.1117/12.856103}

\bibitem[\protect\citeauthoryear{{Batalha} et~al.,}{{Batalha}
  et~al.}{2013}]{2013ApJS..204...24B}
{Batalha} N.~M.,  et~al., 2013, \mn@doi [The Astrophysical Journal Supplement
  Series] {10.1088/0067-0049/204/2/24}, \href
  {https://ui.adsabs.harvard.edu/#abs/2013ApJS..204...24B} {204, 24}

\bibitem[\protect\citeauthoryear{{Boisse} et~al.,}{{Boisse}
  et~al.}{2012}]{2012A&A...545A..55B}
{Boisse} I.,  et~al., 2012, \mn@doi [\aap] {10.1051/0004-6361/201118419}, \href
  {https://ui.adsabs.harvard.edu/#abs/2012A&A...545A..55B} {545, A55}

\bibitem[\protect\citeauthoryear{{Borucki} et~al.,}{{Borucki}
  et~al.}{2010}]{2010Sci...327..977B}
{Borucki} W.~J.,  et~al., 2010, \mn@doi [Science] {10.1126/science.1185402},
  \href {https://ui.adsabs.harvard.edu/#abs/2010Sci...327..977B} {327, 977}

\bibitem[\protect\citeauthoryear{{Bouvier} et~al.,}{{Bouvier}
  et~al.}{2018}]{2018A&A...613A..63B}
{Bouvier} J.,  et~al., 2018, \mn@doi [\aap] {10.1051/0004-6361/201731881},
  \href {https://ui.adsabs.harvard.edu/#abs/2018A&A...613A..63B} {613, A63}

\bibitem[\protect\citeauthoryear{{Buder} et~al.,}{{Buder}
  et~al.}{2018}]{2018MNRAS.478.4513B}
{Buder} S.,  et~al., 2018, \mn@doi [\mnras] {10.1093/mnras/sty1281}, \href
  {https://ui.adsabs.harvard.edu/#abs/2018MNRAS.478.4513B} {478, 4513}

\bibitem[\protect\citeauthoryear{{Cirasuolo}}{{Cirasuolo}}{2016}]{2016ASPC..507..109C}
{Cirasuolo} M.,  2016, in {Skillen} I.,  {Balcells} M.,   {Trager} S.,  eds,
  Vol. 507, Multi-Object Spectroscopy in the Next Decade: Big Questions, Large
  Surveys, and Wide Fields. p.~109

\bibitem[\protect\citeauthoryear{{Contro}, {Horner}, {Wittenmyer}, {Marshall}
  \& {Hinse}}{{Contro} et~al.}{2016}]{2016MNRAS.463..191C}
{Contro} B.,  {Horner} J.,  {Wittenmyer} R.~A.,  {Marshall} J.~P.,   {Hinse}
  T.~C.,  2016, \mn@doi [\mnras] {10.1093/mnras/stw1935}, \href
  {https://ui.adsabs.harvard.edu/#abs/2016MNRAS.463..191C} {463, 191}

\bibitem[\protect\citeauthoryear{{Cutri} \& {et al.}}{{Cutri} \& {et
  al.}}{2014}]{2014yCat.2328....0C}
{Cutri} R.~M.,  {et al.} 2014, VizieR Online Data Catalog, \href
  {https://ui.adsabs.harvard.edu/#abs/2014yCat.2328....0C} {p. II/328}

\bibitem[\protect\citeauthoryear{{De Silva}, {Freeman}, {Bland-Hawthorn},
  {Asplund}  \& {Bessell}}{{De Silva} et~al.}{2007}]{2007AJ....133..694D}
{De Silva} G.~M.,  {Freeman} K.~C.,  {Bland-Hawthorn} J.,  {Asplund} M.,
  {Bessell} M.~S.,  2007, \mn@doi [\aj] {10.1086/510131}, \href
  {https://ui.adsabs.harvard.edu/#abs/2007AJ....133..694D} {133, 694}

\bibitem[\protect\citeauthoryear{{De Silva} et~al.,}{{De Silva}
  et~al.}{2015}]{2015MNRAS.449.2604D}
{De Silva} G.~M.,  et~al., 2015, \mn@doi [\mnras] {10.1093/mnras/stv327}, \href
  {https://ui.adsabs.harvard.edu/#abs/2015MNRAS.449.2604D} {449, 2604}

\bibitem[\protect\citeauthoryear{{Driver}, {Davies}, {Meyer}, {Power},
  {Robotham}, {Baldry}, {Liske}  \& {Norberg}}{{Driver}
  et~al.}{2016}]{2016ASSP...42..205D}
{Driver} S.~P.,  {Davies} L.~J.,  {Meyer} M.,  {Power} C.,  {Robotham}
  A.~S.~G.,  {Baldry} I.~K.,  {Liske} J.,   {Norberg} P.,  2016, in
  {Napolitano} N.~R.,  {Longo} G.,  {Marconi} M.,  {Paolillo} M.,   {Iodice}
  E.,  eds,  Vol. 42, The Universe of Digital Sky Surveys. p.~205,
  \mn@doi{10.1007/978-3-319-19330-4_32}

\bibitem[\protect\citeauthoryear{{Evans} et~al.,}{{Evans}
  et~al.}{2018}]{2018arXiv180409368E}
{Evans} D.~W.,  et~al., 2018, preprint, \href
  {https://ui.adsabs.harvard.edu/#abs/2018arXiv180409368E} {p.
  arXiv:1804.09368} (\mn@eprint {arXiv} {1804.09368})

\bibitem[\protect\citeauthoryear{{Gagn{\'e}} \& {Faherty}}{{Gagn{\'e}} \&
  {Faherty}}{2018}]{2018arXiv180511715G}
{Gagn{\'e}} J.,  {Faherty} J.~K.,  2018, preprint, \href
  {https://ui.adsabs.harvard.edu/#abs/2018arXiv180511715G} {p.
  arXiv:1805.11715} (\mn@eprint {arXiv} {1805.11715})

\bibitem[\protect\citeauthoryear{{Gagn{\'e}} et~al.,}{{Gagn{\'e}}
  et~al.}{2018}]{2018ApJ...856...23G}
{Gagn{\'e}} J.,  et~al., 2018, \mn@doi [\apj] {10.3847/1538-4357/aaae09}, \href
  {https://ui.adsabs.harvard.edu/#abs/2018ApJ...856...23G} {856, 23}

\bibitem[\protect\citeauthoryear{{Gaia Collaboration} et~al.,}{{Gaia
  Collaboration} et~al.}{2016}]{2016A&A...595A...1G}
{Gaia Collaboration} et~al., 2016, \mn@doi [\aap]
  {10.1051/0004-6361/201629272}, \href
  {https://ui.adsabs.harvard.edu/#abs/2016A&A...595A...1G} {595, A1}

\bibitem[\protect\citeauthoryear{{Gaia Collaboration}, {Brown}, {Vallenari},
  {Prusti}, {de Bruijne}, {Babusiaux}  \& {Bailer-Jones}}{{Gaia Collaboration}
  et~al.}{2018}]{2018arXiv180409365G}
{Gaia Collaboration} {Brown} A.~G.~A.,  {Vallenari} A.,  {Prusti} T.,  {de
  Bruijne} J.~H.~J.,  {Babusiaux} C.,   {Bailer-Jones} C.~A.~L.,  2018,
  preprint, \href {https://ui.adsabs.harvard.edu/#abs/2018arXiv180409365G} {p.
  arXiv:1804.09365} (\mn@eprint {arXiv} {1804.09365})

\bibitem[\protect\citeauthoryear{{Gilmore} et~al.,}{{Gilmore}
  et~al.}{2012}]{2012Msngr.147...25G}
{Gilmore} G.,  et~al., 2012, The Messenger, \href
  {https://ui.adsabs.harvard.edu/#abs/2012Msngr.147...25G} {147, 25}

\bibitem[\protect\citeauthoryear{Ginsburg et~al.,}{Ginsburg
  et~al.}{2018}]{astroquery}
Ginsburg A.,  et~al., 2018, astropy/astroquery: v0.3.7 release,
  \mn@doi{10.5281/zenodo.1160627}, \url
  {https://doi.org/10.5281/zenodo.1160627}

\bibitem[\protect\citeauthoryear{{G{\"o}tberg}, {Davies}, {Mustill}, {Johansen}
   \& {Church}}{{G{\"o}tberg} et~al.}{2016}]{2016A&A...592A.147G}
{G{\"o}tberg} Y.,  {Davies} M.~B.,  {Mustill} A.~J.,  {Johansen} A.,   {Church}
  R.~P.,  2016, \mn@doi [\aap] {10.1051/0004-6361/201526309}, \href
  {https://ui.adsabs.harvard.edu/#abs/2016A&A...592A.147G} {592, A147}

\bibitem[\protect\citeauthoryear{{Go{\'z}dziewski} \&
  {Migaszewski}}{{Go{\'z}dziewski} \&
  {Migaszewski}}{2014}]{2014MNRAS.440.3140G}
{Go{\'z}dziewski} K.,  {Migaszewski} C.,  2014, \mn@doi [\mnras]
  {10.1093/mnras/stu455}, \href
  {https://ui.adsabs.harvard.edu/#abs/2014MNRAS.440.3140G} {440, 3140}

\bibitem[\protect\citeauthoryear{{Jeffries}}{{Jeffries}}{2014}]{2014EAS....65..289J}
{Jeffries} R.~D.,  2014, in EAS Publications Series. pp 289--325 (\mn@eprint
  {arXiv} {1404.7156}), \mn@doi{10.1051/eas/1465008}

\bibitem[\protect\citeauthoryear{{Kollmeier} et~al.,}{{Kollmeier}
  et~al.}{2017}]{2017arXiv171103234K}
{Kollmeier} J.~A.,  et~al., 2017, preprint, \href
  {https://ui.adsabs.harvard.edu/#abs/2017arXiv171103234K} {p.
  arXiv:1711.03234} (\mn@eprint {arXiv} {1711.03234})

\bibitem[\protect\citeauthoryear{{Konopacky}, {Marois}, {Macintosh},
  {Galicher}, {Barman}, {Metchev}  \& {Zuckerman}}{{Konopacky}
  et~al.}{2016}]{2016AJ....152...28K}
{Konopacky} Q.~M.,  {Marois} C.,  {Macintosh} B.~A.,  {Galicher} R.,  {Barman}
  T.~S.,  {Metchev} S.~A.,   {Zuckerman} B.,  2016, \mn@doi [\aj]
  {10.3847/0004-6256/152/2/28}, \href
  {https://ui.adsabs.harvard.edu/#abs/2016AJ....152...28K} {152, 28}

\bibitem[\protect\citeauthoryear{{Kos} et~al.,}{{Kos}
  et~al.}{2017}]{2017MNRAS.464.1259K}
{Kos} J.,  et~al., 2017, \mn@doi [\mnras] {10.1093/mnras/stw2064}, \href
  {https://ui.adsabs.harvard.edu/#abs/2017MNRAS.464.1259K} {464, 1259}

\bibitem[\protect\citeauthoryear{{Kraus}, {Herczeg}, {Rizzuto}, {Mann},
  {Slesnick}, {Carpenter}, {Hillenbrand}  \& {Mamajek}}{{Kraus}
  et~al.}{2017}]{2017ApJ...838..150K}
{Kraus} A.~L.,  {Herczeg} G.~J.,  {Rizzuto} A.~C.,  {Mann} A.~W.,  {Slesnick}
  C.~L.,  {Carpenter} J.~M.,  {Hillenbrand} L.~A.,   {Mamajek} E.~E.,  2017,
  \mn@doi [\apj] {10.3847/1538-4357/aa62a0}, \href
  {https://ui.adsabs.harvard.edu/#abs/2017ApJ...838..150K} {838, 150}

\bibitem[\protect\citeauthoryear{{Kunder} et~al.,}{{Kunder}
  et~al.}{2017}]{2017AJ....153...75K}
{Kunder} A.,  et~al., 2017, \mn@doi [\aj] {10.3847/1538-3881/153/2/75}, \href
  {https://ui.adsabs.harvard.edu/#abs/2017AJ....153...75K} {153, 75}

\bibitem[\protect\citeauthoryear{{Lind}, {Asplund}  \& {Barklem}}{{Lind}
  et~al.}{2009}]{2009A&A...503..541L}
{Lind} K.,  {Asplund} M.,   {Barklem} P.~S.,  2009, \mn@doi [\aap]
  {10.1051/0004-6361/200912221}, \href
  {https://ui.adsabs.harvard.edu/#abs/2009A&A...503..541L} {503, 541}

\bibitem[\protect\citeauthoryear{{Luo} et~al.,}{{Luo}
  et~al.}{2015}]{2015RAA....15.1095L}
{Luo} A.~L.,  et~al., 2015, \mn@doi [Research in Astronomy and Astrophysics]
  {10.1088/1674-4527/15/8/002}, \href
  {https://ui.adsabs.harvard.edu/#abs/2015RAA....15.1095L} {15, 1095}

\bibitem[\protect\citeauthoryear{{Majewski} et~al.,}{{Majewski}
  et~al.}{2017}]{2017AJ....154...94M}
{Majewski} S.~R.,  et~al., 2017, \mn@doi [\aj] {10.3847/1538-3881/aa784d},
  \href {https://ui.adsabs.harvard.edu/#abs/2017AJ....154...94M} {154, 94}

\bibitem[\protect\citeauthoryear{{Mamajek} \& {Hillenbrand}}{{Mamajek} \&
  {Hillenbrand}}{2008}]{2008ApJ...687.1264M}
{Mamajek} E.~E.,  {Hillenbrand} L.~A.,  2008, \mn@doi [\apj] {10.1086/591785},
  \href {https://ui.adsabs.harvard.edu/#abs/2008ApJ...687.1264M} {687, 1264}

\bibitem[\protect\citeauthoryear{{Marcy}, {Butler}, {Fischer}, {Laughlin},
  {Vogt}, {Henry}  \& {Pourbaix}}{{Marcy} et~al.}{2002}]{2002ApJ...581.1375M}
{Marcy} G.~W.,  {Butler} R.~P.,  {Fischer} D.~A.,  {Laughlin} G.,  {Vogt}
  S.~S.,  {Henry} G.~W.,   {Pourbaix} D.,  2002, \mn@doi [\apj]
  {10.1086/344298}, \href
  {https://ui.adsabs.harvard.edu/#abs/2002ApJ...581.1375M} {581, 1375}

\bibitem[\protect\citeauthoryear{{Marois}, {Macintosh}, {Barman}, {Zuckerman},
  {Song}, {Patience}, {Lafreni{\`e}re}  \& {Doyon}}{{Marois}
  et~al.}{2008}]{2008Sci...322.1348M}
{Marois} C.,  {Macintosh} B.,  {Barman} T.,  {Zuckerman} B.,  {Song} I.,
  {Patience} J.,  {Lafreni{\`e}re} D.,   {Doyon} R.,  2008, \mn@doi [Science]
  {10.1126/science.1166585}, \href
  {https://ui.adsabs.harvard.edu/#abs/2008Sci...322.1348M} {322, 1348}

\bibitem[\protect\citeauthoryear{{Marois}, {Zuckerman}, {Konopacky},
  {Macintosh}  \& {Barman}}{{Marois} et~al.}{2010}]{2010Natur.468.1080M}
{Marois} C.,  {Zuckerman} B.,  {Konopacky} Q.~M.,  {Macintosh} B.,   {Barman}
  T.,  2010, \mn@doi [\nat] {10.1038/nature09684}, \href
  {https://ui.adsabs.harvard.edu/#abs/2010Natur.468.1080M} {468, 1080}

\bibitem[\protect\citeauthoryear{{Marshall}, {Horner}  \& {Carter}}{{Marshall}
  et~al.}{2010}]{2010IJAsB...9..259M}
{Marshall} J.,  {Horner} J.,   {Carter} A.,  2010, \mn@doi [International
  Journal of Astrobiology] {10.1017/S1473550410000297}, \href
  {https://ui.adsabs.harvard.edu/#abs/2010IJAsB...9..259M} {9, 259}

\bibitem[\protect\citeauthoryear{{Martell} et~al.,}{{Martell}
  et~al.}{2017}]{2017MNRAS.465.3203M}
{Martell} S.~L.,  et~al., 2017, \mn@doi [\mnras] {10.1093/mnras/stw2835}, \href
  {https://ui.adsabs.harvard.edu/#abs/2017MNRAS.465.3203M} {465, 3203}

\bibitem[\protect\citeauthoryear{{Mentuch}, {Brandeker}, {van Kerkwijk},
  {Jayawardhana}  \& {Hauschildt}}{{Mentuch}
  et~al.}{2008}]{2008ApJ...689.1127M}
{Mentuch} E.,  {Brandeker} A.,  {van Kerkwijk} M.~H.,  {Jayawardhana} R.,
  {Hauschildt} P.~H.,  2008, \mn@doi [\apj] {10.1086/592764}, \href
  {https://ui.adsabs.harvard.edu/#abs/2008ApJ...689.1127M} {689, 1127}

\bibitem[\protect\citeauthoryear{{Mullally} et~al.,}{{Mullally}
  et~al.}{2015}]{2015ApJS..217...31M}
{Mullally} F.,  et~al., 2015, \mn@doi [The Astrophysical Journal Supplement
  Series] {10.1088/0067-0049/217/2/31}, \href
  {https://ui.adsabs.harvard.edu/#abs/2015ApJS..217...31M} {217, 31}

\bibitem[\protect\citeauthoryear{{Noyes}, {Hartmann}, {Baliunas}, {Duncan}  \&
  {Vaughan}}{{Noyes} et~al.}{1984}]{1984ApJ...279..763N}
{Noyes} R.~W.,  {Hartmann} L.~W.,  {Baliunas} S.~L.,  {Duncan} D.~K.,
  {Vaughan} A.~H.,  1984, \mn@doi [\apj] {10.1086/161945}, \href
  {https://ui.adsabs.harvard.edu/#abs/1984ApJ...279..763N} {279, 763}

\bibitem[\protect\citeauthoryear{{Pavlenko} \& {Magazzu}}{{Pavlenko} \&
  {Magazzu}}{1996}]{1996A&A...311..961P}
{Pavlenko} Y.~V.,  {Magazzu} A.,  1996, \aap, \href
  {https://ui.adsabs.harvard.edu/#abs/1996A&A...311..961P} {311, 961}

\bibitem[\protect\citeauthoryear{{Pecaut} \& {Mamajek}}{{Pecaut} \&
  {Mamajek}}{2013}]{2013ApJS..208....9P}
{Pecaut} M.~J.,  {Mamajek} E.~E.,  2013, \mn@doi [\apjs]
  {10.1088/0067-0049/208/1/9}, \href
  {http://adsabs.harvard.edu/abs/2013ApJS..208....9P} {208, 9}

\bibitem[\protect\citeauthoryear{{Perryman} et~al.,}{{Perryman}
  et~al.}{1998}]{1998A&A...331...81P}
{Perryman} M.~A.~C.,  et~al., 1998, \aap, \href
  {https://ui.adsabs.harvard.edu/#abs/1998A&A...331...81P} {331, 81}

\bibitem[\protect\citeauthoryear{{Perryman}, {Hartman}, {Bakos}  \&
  {Lindegren}}{{Perryman} et~al.}{2014}]{2014ApJ...797...14P}
{Perryman} M.,  {Hartman} J.,  {Bakos} G.~{\'A}.,   {Lindegren} L.,  2014,
  \mn@doi [\apj] {10.1088/0004-637X/797/1/14}, \href
  {https://ui.adsabs.harvard.edu/#abs/2014ApJ...797...14P} {797, 14}

\bibitem[\protect\citeauthoryear{{Ricker} et~al.,}{{Ricker}
  et~al.}{2015}]{2015JATIS...1a4003R}
{Ricker} G.~R.,  et~al., 2015, \mn@doi [Journal of Astronomical Telescopes,
  Instruments, and Systems] {10.1117/1.JATIS.1.1.014003}, \href
  {https://ui.adsabs.harvard.edu/#abs/2015JATIS...1a4003R} {1, 014003}

\bibitem[\protect\citeauthoryear{{Rizzuto}, {Mann}, {Vanderburg}, {Kraus}  \&
  {Covey}}{{Rizzuto} et~al.}{2017}]{2017AJ....154..224R}
{Rizzuto} A.~C.,  {Mann} A.~W.,  {Vanderburg} A.,  {Kraus} A.~L.,   {Covey}
  K.~R.,  2017, \mn@doi [\aj] {10.3847/1538-3881/aa9070}, \href
  {https://ui.adsabs.harvard.edu/#abs/2017AJ....154..224R} {154, 224}

\bibitem[\protect\citeauthoryear{{Sartoretti} et~al.,}{{Sartoretti}
  et~al.}{2018}]{2018A&A...616A...6S}
{Sartoretti} P.,  et~al., 2018, \mn@doi [\aap] {10.1051/0004-6361/201832836},
  \href {https://ui.adsabs.harvard.edu/#abs/2018A&A...616A...6S} {616, A6}

\bibitem[\protect\citeauthoryear{{Schmidt}, {Hawley}, {West}, {Bochanski},
  {Davenport}, {Ge}  \& {Schneider}}{{Schmidt}
  et~al.}{2015}]{2015AJ....149..158S}
{Schmidt} S.~J.,  {Hawley} S.~L.,  {West} A.~A.,  {Bochanski} J.~J.,
  {Davenport} J. R.~A.,  {Ge} J.,   {Schneider} D.~P.,  2015, \mn@doi [\aj]
  {10.1088/0004-6256/149/5/158}, \href
  {https://ui.adsabs.harvard.edu/#abs/2015AJ....149..158S} {149, 158}

\bibitem[\protect\citeauthoryear{{Skrutskie} et~al.,}{{Skrutskie}
  et~al.}{2006}]{2006AJ....131.1163S}
{Skrutskie} M.~F.,  et~al., 2006, \mn@doi [\aj] {10.1086/498708}, \href
  {https://ui.adsabs.harvard.edu/#abs/2006AJ....131.1163S} {131, 1163}

\bibitem[\protect\citeauthoryear{{Snellen} \& {Brown}}{{Snellen} \&
  {Brown}}{2018}]{2018arXiv180806257S}
{Snellen} I.,  {Brown} A.,  2018, preprint, \href
  {https://ui.adsabs.harvard.edu/#abs/2018arXiv180806257S} {p.
  arXiv:1808.06257} (\mn@eprint {arXiv} {1808.06257})

\bibitem[\protect\citeauthoryear{{Soderblom}}{{Soderblom}}{2010}]{2010ARA&A..48..581S}
{Soderblom} D.~R.,  2010, \mn@doi [Annual Review of Astronomy and Astrophysics]
  {10.1146/annurev-astro-081309-130806}, \href
  {https://ui.adsabs.harvard.edu/#abs/2010ARA&A..48..581S} {48, 581}

\bibitem[\protect\citeauthoryear{{Soderblom}, {Hillenbrand}, {Jeffries},
  {Mamajek}  \& {Naylor}}{{Soderblom} et~al.}{2014}]{2014prpl.conf..219S}
{Soderblom} D.~R.,  {Hillenbrand} L.~A.,  {Jeffries} R.~D.,  {Mamajek} E.~E.,
  {Naylor} T.,  2014, in Protostars and Planets VI. p.~219 (\mn@eprint {arXiv}
  {1311.7024}), \mn@doi{10.2458/azu_uapress_9780816531240-ch010}

\bibitem[\protect\citeauthoryear{{Traven} et~al.,}{{Traven}
  et~al.}{2017}]{2017ApJS..228...24T}
{Traven} G.,  et~al., 2017, \mn@doi [The Astrophysical Journal Supplement
  Series] {10.3847/1538-4365/228/2/24}, \href
  {https://ui.adsabs.harvard.edu/#abs/2017ApJS..228...24T} {228, 24}

\bibitem[\protect\citeauthoryear{{Wertz}, {Absil}, {G{\'o}mez Gonz{\'a}lez},
  {Milli}, {Girard}, {Mawet}  \& {Pueyo}}{{Wertz}
  et~al.}{2017}]{2017A&A...598A..83W}
{Wertz} O.,  {Absil} O.,  {G{\'o}mez Gonz{\'a}lez} C.~A.,  {Milli} J.,
  {Girard} J.~H.,  {Mawet} D.,   {Pueyo} L.,  2017, \mn@doi [\aap]
  {10.1051/0004-6361/201628730}, \href
  {https://ui.adsabs.harvard.edu/#abs/2017A&A...598A..83W} {598, A83}

\bibitem[\protect\citeauthoryear{{West}, {Hawley}, {Bochanski}, {Covey},
  {Reid}, {Dhital}, {Hilton}  \& {Masuda}}{{West}
  et~al.}{2008}]{2008AJ....135..785W}
{West} A.~A.,  {Hawley} S.~L.,  {Bochanski} J.~J.,  {Covey} K.~R.,  {Reid}
  I.~N.,  {Dhital} S.,  {Hilton} E.~J.,   {Masuda} M.,  2008, \mn@doi [\aj]
  {10.1088/0004-6256/135/3/785}, \href
  {https://ui.adsabs.harvard.edu/#abs/2008AJ....135..785W} {135, 785}

\bibitem[\protect\citeauthoryear{{Wittenmyer} et~al.,}{{Wittenmyer}
  et~al.}{2014}]{2014ApJ...783..103W}
{Wittenmyer} R.~A.,  et~al., 2014, \mn@doi [\apj]
  {10.1088/0004-637X/783/2/103}, \href
  {https://ui.adsabs.harvard.edu/#abs/2014ApJ...783..103W} {783, 103}

\bibitem[\protect\citeauthoryear{{Wittenmyer} et~al.,}{{Wittenmyer}
  et~al.}{2016}]{2016ApJ...819...28W}
{Wittenmyer} R.~A.,  et~al., 2016, \mn@doi [\apj] {10.3847/0004-637X/819/1/28},
  \href {https://ui.adsabs.harvard.edu/#abs/2016ApJ...819...28W} {819, 28}

\bibitem[\protect\citeauthoryear{{Yanny} et~al.,}{{Yanny}
  et~al.}{2009}]{2009AJ....137.4377Y}
{Yanny} B.,  et~al., 2009, \mn@doi [\aj] {10.1088/0004-6256/137/5/4377}, \href
  {https://ui.adsabs.harvard.edu/#abs/2009AJ....137.4377Y} {137, 4377}

\bibitem[\protect\citeauthoryear{{Zhao} et~al.,}{{Zhao}
  et~al.}{2015}]{2015RAA....15.1282Z}
{Zhao} J.-K.,  et~al., 2015, \mn@doi [Research in Astronomy and Astrophysics]
  {10.1088/1674-4527/15/8/013}, \href
  {https://ui.adsabs.harvard.edu/#abs/2015RAA....15.1282Z} {15, 1282}

\bibitem[\protect\citeauthoryear{{Zwitter} et~al.,}{{Zwitter}
  et~al.}{2018}]{2018MNRAS.481..645Z}
{Zwitter} T.,  et~al., 2018, \mn@doi [\mnras] {10.1093/mnras/sty2293}, \href
  {https://ui.adsabs.harvard.edu/#abs/2018MNRAS.481..645Z} {481, 645}

\bibitem[\protect\citeauthoryear{{de Jong} et~al.,}{{de Jong}
  et~al.}{2012}]{2012SPIE.8446E..0TD}
{de Jong} R.~S.,  et~al., 2012, in Ground-based and Airborne Instrumentation
  for Astronomy IV. p. 84460T (\mn@eprint {arXiv} {1206.6885}),
  \mn@doi{10.1117/12.926239}

\bibitem[\protect\citeauthoryear{{de Laverny}, {Recio-Blanco}, {Worley}  \&
  {Plez}}{{de Laverny} et~al.}{2012}]{2012A&A...544A.126D}
{de Laverny} P.,  {Recio-Blanco} A.,  {Worley} C.~C.,   {Plez} B.,  2012,
  \mn@doi [\aap] {10.1051/0004-6361/201219330}, \href
  {https://ui.adsabs.harvard.edu/#abs/2012A&A...544A.126D} {544, A126}

\bibitem[\protect\citeauthoryear{{{\v{Z}}erjal} et~al.,}{{{\v{Z}}erjal}
  et~al.}{2017}]{2017ApJ...835...61Z}
{{\v{Z}}erjal} M.,  et~al., 2017, \mn@doi [\apj] {10.3847/1538-4357/835/1/61},
  \href {https://ui.adsabs.harvard.edu/#abs/2017ApJ...835...61Z} {835, 61}

\makeatother
\end{thebibliography}

\bsp	
\label{lastpage}
\end{document}